\documentstyle[epsf,preprint,floats,prd,aps]{revtex}
\newif\iftwocolumn
\def\figfmh{
\begin{figure}
\iftwocolumn\epsfxsize=\columnwidth\else\epsfysize=5in\fi
\begin{center}
\leavevmode
\epsfbox[100 330 500 758]{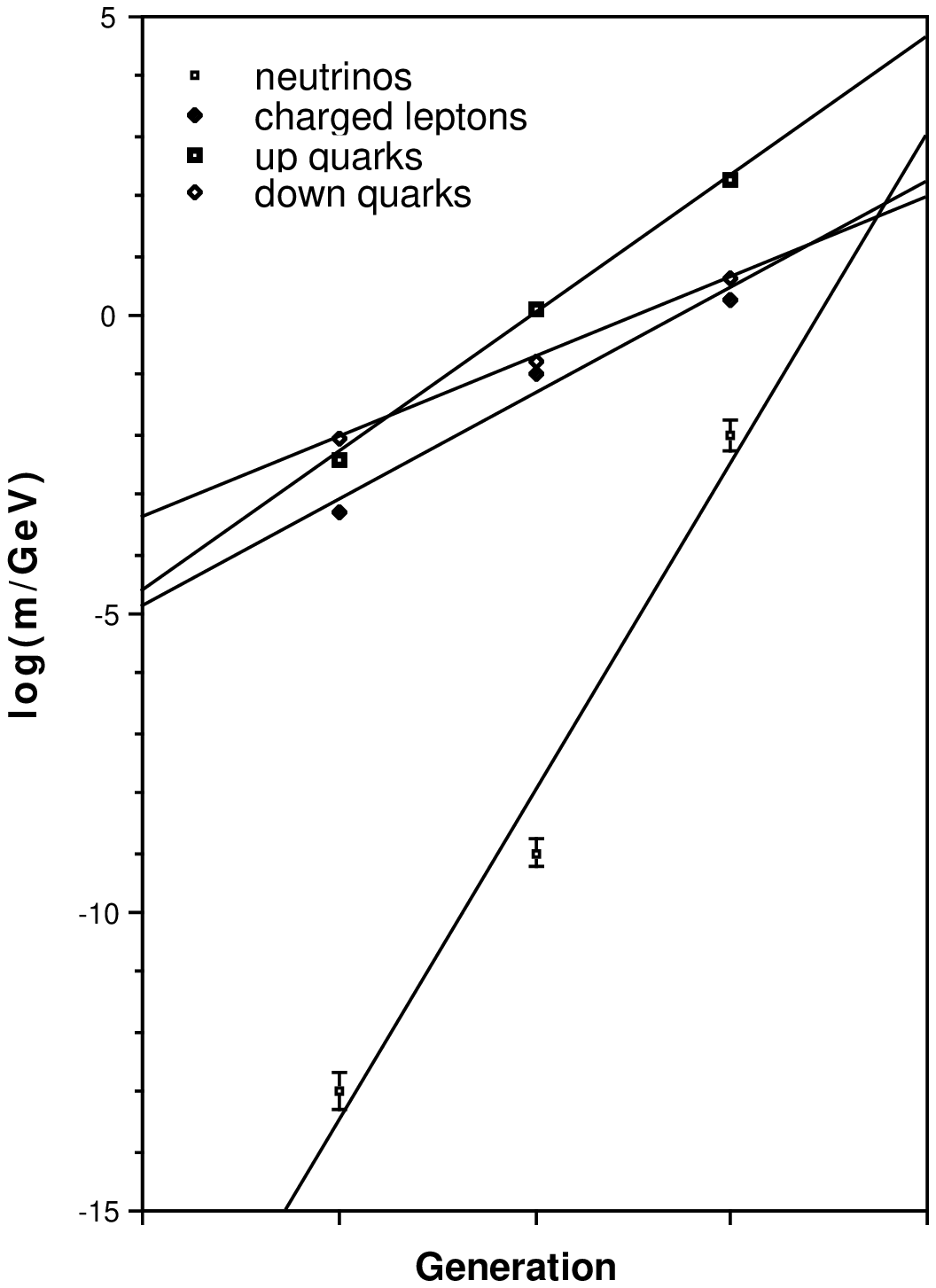}
\end{center}
\caption[Fermion mass hierarchies]{Fermion mass hierarchies. The up and
down quark and charged lepton hierarchies are shown. We include the proposed
Dirac neutrino masses. The bars indicate the range of masses consistent
with experiments.}
\label{hierarchy}
\end{figure}
}
\def\etal{{\it et al.}}
\def\diag{{\rm diag}}
\def\sci#1#2{#1\times10^{#2}}
\def\dms{\delta M^2}
\def\sstt{\mbox{$\sin^2 2\theta$}}
\def\ssd#1{\sin^2 {\frac{\Delta_{#1}}{2}}}
\def\Sm#1{\O{10^{-#1}}}
\def\LD{{\cal L}}
\def\e{\epsilon}
\def\nubar{\bar\nu}
\def\ebar{\bar e}

\def\eV{{\rm eV}}
\def\meV{{\rm meV}}
\def\O#1{{\cal O}(#1)}
\def\Re{{\rm Re}}

\begin{document}

\draft

\preprint{\vbox{
\hbox{CTP-TAMU-48/96}
\hbox{hep-ph/9609509}
\hbox{September 1996}
}}

\title{Neutrino Oscillations from Dirac and Majorana Masses}

\author{D.~Ring\thanks{Email address: Cdude@phys.tamu.edu}}

\address{Center for Theoretical Physics, Department of Physics,
		Texas A\&M University, 
  		College Station, TX 77843-4242}

\maketitle

\begin{abstract}

We present a scenario of neutrino masses and mixing angles.
Each generation includes a sterile right handed
neutrino in addition to the usual left handed one. We assume a hierarchy in
their Dirac masses similar to, but much larger than the hierarchies in the
quarks and charged leptons. In addition, we include a Majorana mass term
for the sterile neutrinos only. These assumptions prove
sufficient to accomodate scales of mass differences and mixing angles
consistent with all existing neutrino oscillation data.

\end{abstract}

\pacs{}

\section{Introduction}

While there is currently no direct experimental evidence for neutrino masses,
there is growing indirect evidence in the form of neutrino oscillations,
culminating in the recent observation of
$\overline\nu_\mu\rightarrow\overline\nu_e$ at LSND\cite{LSND}.
The combined evidence suggests three independent mass splittings among
the neutrinos participating in the oscillations. If each of these splittings
is taken seriously, then a fourth neutrino is required to accomodate
all the data\cite{need/sterile}.
Several such scenarios have been proposed\cite{4/nu}. 
In most cases small mass differences and small mixing angles are put in
by hand, and it seems difficult to explain their origin without fine tuning.
Furthermore, these scenarios treat generations on an unequal footing,
mixing the extra sterile state with only the electron neutrino.

We examine here the viability of one sterile neutrino for each generation.
Such models have not been considered previously due to the constraint
$N_\nu \lesssim 4$ from big bang nucleosynthesis\cite{BBN/Limit}. 
It is known, however, that this constraint can be avoided if the tau
neutrinos have masses in the MeV range and decay rapidly into
$\nu_e$\cite{tau/MeV}. The three mass splittings then suggest a
unique natural mass spectrum. When combined with various terrestrial
experimental data, the solar neutrino deficit implies a neutrino almost
degenerate with $\nu_e$, the atmospheric deficit implies a
neutrino almost degenerate with $\nu_\mu$, and the LSND data
implies the two pairs must be split by at least $0.1\,\eV$.
We therefore impose Dirac masses with a very large hierarchy
and a CKM matrix analogous to the one in the quark sector. We then include
a Majorana mass matrix on the right handed neutrinos only. The scale of the 
Majorana masses is $\O{10^{-2}\,\eV}$ and is appropriate for a seesaw
mechanism\cite{seesaw} between the
Grand Unification (GUT) scale and the electroweak scale. 
Neutrino mixings appropriate for the oscillation data will arise
from the interplay between the Dirac and Majorana mass matrices.

\section{Fermion mass hierarchies}

Typically a model with
a large mass hierarchy will have mass matrix elements whose scales obey
\begin{equation}
m_{ij} \lesssim
\left( \begin{array}{ccc}
\e^4&\e^3&\e^2\\
\e^3&\e^2&\e\\
\e^2&\e&1\\
\end{array} \right)\times m_0; \qquad\qquad \e \ll 1\,.
\label{Md}
\end{equation}
Here $m_0$ is the scale of the largest mass eigenvalue.
If the off diagonal elements are larger than these, the lighter masses
will recieve large seesaw contributions and the hierarchy will be destroyed.
Note that these are only upper bounds on the scales of the matrix elements.
Realistic models typically contain texture
zeros\cite{textures} or additional powers of $\e$ in their off
diagonal elements. Such suppressions are in fact necessary to make
contact with the Standard Model, as we will see below.
In the Standard Model we have $\e_u\sim1/14.3$, $\e_d\sim1/5.1$,
and $\e_e\sim1/7.6$ (see Figure \ref{hierarchy}).
With $m_{\nu_\tau}$ in the MeV range, the
effects of the off diagonal elements of the neutrino Dirac mass matrix
will be washed out by the charged lepton mixings.

The Dirac mass term in the neutrino Lagrangian is
\begin{equation}
\LD = -\bar\nu'_i m_{ij} \nu'_j = -\bar\nu'_{L_i} m_{ij} \nu'_{R_j} + H.c.
\label{Dirac}
\end{equation}
$m_{ij}$ is diagonalized by
\begin{eqnarray}
\nubar_L''&= \nubar_L' \, {U_L^\nu}^\dagger \nonumber \\
\nu_R&= U_R^\nu \nu'_R \label{diag}
\end{eqnarray}
and the charged leptons are diagonalized by
\begin{eqnarray}
\ebar_L&= \ebar'_L \, {U_L^e}^\dagger \nonumber \\
e_R&= U_R^e \, e'_R \, ,
\end{eqnarray}
where we have suppressed generation indices.
Thus for example, the neutrino mass term can be written
\begin{equation}
\LD_\nu =
\nubar''_{L_i} U_L^\nu m_{ij} {U_R^\nu}^\dagger \nu_{R_j} + H.c.
\label{nmt}
\end{equation}
where $U_L^\nu m_{ij} {U_R^\nu}^\dagger = \diag(m_1,m_2,m_3)$.

If both the up-like and down-like members of an $SU(2)$ multiplet have
mass matrices like (\ref{Md}), then the down-like mass eigenstates will
be rotated from the weak partners of the up-like mass states by the CKM
matrix,
\begin{equation}
V = U^u_L {U^d_L}^\dagger \sim
\left( \begin{array}{ccc}
1&\e&\e^2\\
\e&1&\e\\
\e^2&\e&1
\end{array} \right),
\label{Vckm}
\end{equation}
where $\e$ is the larger of $\e_u$ and $\e_d$. For example, the weak partner
of the electron would be
\begin{equation}
\nu_L = U_L^e \nu'_L = V^l \nu''_L
\label{nckm}
\end{equation}
and the CKM hierarchy parameter is $\e_e$.
We assume the hierarchies
of the $e$ and $\nu$ states
are aligned with respect to each other. It is possible to consider two
hierarchies which are related by a
generic unitary transformation, but it is hard to imagine a mechanism which
would generate such hierarchies naturally. In a generic basis, the
matrix elements would appear to be fine tuned to $\O{\e^4}$. 
The alignment of the quark hierarchies
is evident in the smallness of the off diagonal elements of their CKM matrix.
The Cabibbo angle in particular is very close to its expected magnitude.
Note, however, that the other off diagonal elements of $V^q$ are
smaller by an additional factor of $\e_d$ than would be expected on the
basis of the quark hierarchies alone. Thus $V^q$ is well described by
the Wolfenstein parametrization\cite{Wolf/Param},
\begin{equation}
V^q =
\left( \begin{array}{ccc}
1-\e^2/2 &	\e &		\e^3 A \bar z\\
-\e &		1-\e^2/2 &	\e^2A\\
\e^3 A (1-z) &	-\e^2A &	1
\end{array} \right),
\label{Wolf}
\end{equation}
where $A \sim \O{1}$ is real, $\e \sim \e_d$ is real, and $z$ is a complex
number with magnitude $\O{1}$.
{\it A~priori} there is no way to know whether $V^l$ will follow this
pattern or the pattern of (\ref{Vckm}) or some other texture.
For definiteness we first use pattern (\ref{Wolf}) with $\e \sim \e_e$ as
an example and then discuss other possibilities.

\section{The Majorana mass matrix}

In order to achieve mass splittings and mixing angles appropriate for
the Atmospheric and Solar oscillation data, we include Majorana masses
of $\O{10^{-2}\,\eV}$ for the right handed
neutrinos. This is about the right scale to be generated by a seesaw between
the GUT scale and the electroweak scale, although it is not obvious how
sterile particles relate to the electroweak scale. It is interesting to
note that with this interpretation, the electron neutrino will be at the
bottom of a two stage seesaw. We write the Majorana mass matrix as
\begin{equation}
m_{ij}=
\left( \begin{array}{ccc}
a_{11}&a_{12}&a_{13}\\
a_{12}&a_{22}&a_{23}\\
a_{13}&a_{23}&a_{33}
\end{array} \right),
\label{Mm}
\end{equation}
where {\it a priori} each $a_{ij}$ is $\O{10^{-2}\,\eV}$.

We are now in a position to write the full
$6\times6$ mass matrix. We write the right handed neutrinos in terms of their
charge conjugates, $\nu_R = s^c\,$. We use (\ref{Mm}), (\ref{nmt}),
and (\ref{nckm}) with the pattern (\ref{Wolf}) to get
\begin{equation}
\LD_\nu = {1\over2} N C M' N + H.c.
\end{equation}
where $C$ is the charge conjugation matrix,
$N=(\nu_e, s_e, \nu_\mu, s_\mu, \nu_\tau, s_\tau)$, and
\begin{equation}
\arraycolsep-1pt
M' =
\left( \begin{array}{cccccc}
0&  m_1(1-\e^2/2)&  0&  m_2\e&  0&  m_3\e^3Az\\
m_1(1-\e^2/2)&  a_{11}&  -m_1\e&  a_{12}&  m_1\e^3A(1-{\bar z})&  a_{13}\\
0&  -m_1\e&  0&  m_2(1-\e^2/2)&  0&  m_3\e^2A\\
m_2\e&  a_{12}&  m_2(1-\e^2/2)&  a_{22}&  -m_2\e^2A&  a_{23}\\
0&  m_1\e^3A(1-{\bar z})&  0&  -m_2\e^2A&  0&  m_3\\
m_3\e^3Az&  a_{13}&  m_3\e^2A&  a_{23}&  m_3&  a_{33}\\
\end{array} \right).
\label{M6}
\end{equation}
The $m_i$ are defined by (\ref{nmt}).

We take ${\rm Im}(z) = 0$. We will not consider CP violation here.
$M'$ is diagonalized by $M' = O P M P \tilde O$, where $O$ is orthogonal
and $P$ is a diagonal phase matrix. The weak eigenstates are written
in terms of the mass eigenstates by
$\nu_\alpha~=~U_{\alpha i}~\nu_i~=~O_{i\alpha}P_i~\nu_i$. We have
\begin{equation}
\begin{array}{rl}
\nu_1=&i(\,1-{\e^2\over2}\,,\,-{m_1\over a_{11}}\,,\,-\e\,,\,
\Sm{10}\,,\, -\e^3A(z-1)\,,\, \Sm{19} ) \\
\nu_2=&(\,{m_1\over a_{11}}-{\e a_{12}\over m_2}\,,\,
1-{m_1^2\over2a^2_{11}}-{a^2_{12}\over 2m_2^2}\,,\,
-{a_{12}\over m_2}-{\e m_1\over a_{11}}\,,\,-{a_{11}a_{12}\over m_2^2}\,,\,
{\e^2 A a_{12}\over m_2}\,,\, \Sm{15} ) \\
\nu_3=&{i\over\sqrt{2}}(\,\e\,,\,-{a_{12}\over m_2}\,,\,
1-{\e^2\over2}+{a_{22}\over 4m_2}\,,\,-1+{a_{22}\over4m_2}\,,\,
-\e^2A\,,\, {\e^4Am_2\over m_3}(z-1/2) ) \\
\nu_4=&{1\over\sqrt{2}}(\,\e\,,\,{a_{12}\over m_2}\,,\,
1-{\e^2\over2}-{a_{22}\over4m_2}\,,\, 1+{a_{22}\over4m_2}\,,\,
-\e^2A\,,\, -{\e^4Am_2\over m_3}(z-1/2) ) \\
\nu_5=&{i\over\sqrt{2}}	(\,\e^3Az\,,\,{a_{13}\over m_3}\,,\,\e^2A\,,\,
{a_{23}\over m_3}\,,\,(1-{\e^4A^2\over2})\,,\,-1) \\
\nu_6=&{1\over\sqrt{2}}	(\,\e^3Az\,,\,{a_{13}\over m_3}\,,\,\e^2A\,,\,
{a_{23}\over m_3}\,,\,(1-{\e^4A^2\over2})\,,\,1) \\
\end{array}
\end{equation}
and the masses are
\begin{equation}
M_i = \left( \,{m_1^2\over a_{11}}\,,\, a_{11}\,,\, m_2-{a_{22}\over2}\,,\,
m_2+{a_{22}\over2}\,,\, m_3(1+{\e^4A^2\over2})\,,\, m_3(1+{\e^4A^2\over2})
\right),
\end{equation}
where we have dropped higher order terms.

\section{Experiments}

The probability that an initial $\nu_\alpha$ of energy $E$ will oscillate
into $\nu_\beta$ after a distance $L$ is
\begin{equation}
P(\nu_\alpha \rightarrow \nu_\beta) = \delta_{\alpha\beta} -
2 \sum_{i \ne j} \Re[U_{\alpha i}U^*_{\alpha j}U^*_{\beta i}U_{\beta j}]
\sin^2{\frac{\Delta_{ij}}{2}}
\end{equation}
\begin{equation}
\Delta_{ij} = \frac{L(m_i^2-m_j^2)}{2E} .
\end{equation}

For the LSND experiment, the relevant terms are
\begin{equation}
P_{\mu e} = \e^2(2\ssd{13}+2\ssd{14}-\ssd{34}).
\end{equation}
The last term vanishes since its wavelength is too long for the 30m LSND
baseline. The other terms
are effectively equal, and together have the same effect as a two flavor
$\nu_\mu - \nu_e$ oscillation with $\dms = m_2^2$ and $\sstt = 4\e^2$.
We may therefore appeal to published two flavor analyses.

The LSND experiment was designed to be most
sensitive at the mass splitting preferred by the CHDM model of
cosmological structure formation\cite{CHDM}, $\dms\sim6\,\eV^2$. The
allowed regions of $\sstt$ at that $\dms$ depend on how the data
is analyzed. The
99\% likelihood region is $0.002<\sstt\lesssim0.01$, while
the 80\% {\it confidence level\,} band is $0.0012<\sstt\lesssim0.005$ .
The confidence band uses only the number of events, while the likelihood
region uses all the information about the events including neutrino
energy and distance from source to detection point so it is the best way
to determine favored regions of $\dms$ and $\sstt$\cite{Caldwell/FAX}.
The difference is important because $0.002<\sstt$ is excluded at
90\% confidence level by the BNL E776 experiment\cite{E776}.
Thus if we compare similar types of bounds, there is a marginally allowed
region consistent with the cosmologically preferred $\dms$.

On the other hand, there is a large region at lower $\dms$ allowed
by all the data (including limits from r-process
nucleosynthesis\cite{Fuller/Primack/Qian}). The $90\%$
likelihood region of LSND, combined with the $90\%$ confidence limits
from E776 and the Bugey reactor experiment\cite{Bugey} 
allow $0.25<\dms<2.3\,\eV^2$ and $0.002<\sstt<0.04$,
giving $0.5<m_2<1.5\,\eV$
and $\frac{1}{45}<\e<\frac{1}{10}$. The upper end of the range of $\e$ is
reasonably close to our expectation of $\e \sim \frac{1}{7.6}$ . If
\sstt is found to be in the lower end of this range, then
the mixing of the charged leptons must be supressed from its
expected hierarchical value for almost any conceivable neutrino mass
scenario.

For the atmospheric deficit, the relevant probability is
\begin{equation}
P_{\mu \mu} = 1-\ssd{34}.
\end{equation}
This has the same effect as a two flavor $\nu_\mu - \nu_s$ oscillation with
$\dms = 2 m_2 a_{22}$ and $\sstt = 1$. Maximal mixing is allowed by the
combined Frejus, NUSEX, IMB, Kamioka sub-GeV, and Kamioka multi-GeV zenith
angle dependent data for
$\sci{4}{-4}<\dms<0.01\,\eV^2$\cite{Atmospheric/3/Flavor},
giving $0.13<a_{22}<10\,\meV$ (milli-eV). There is a small probability
$(\sim 2\%)$ of oscillation into $\nu_e$. Matter effects are insignificant.

The vacuum disappearance probability for $\nu_e$ is
\begin{equation}
P^{vac}_{ee} = 1 - 2\e^2\ssd{13} - 2\e^2\ssd{14} .
\end{equation}
This is too small an effect for the solar neutrino deficit, so we assume a
$\nu_e - \nu_s$ small angle MSW mechanism. The allowed parameters
are $0.003<\sstt<0.012$ and $4<\dms<11\,\meV^2$.
We have $\dms=a_{11}^2$ and $\sin \theta = \frac{m_1}{a_{11}}$, giving
$2<a_{11}<3.3\,\meV$
and $0.05<m_1<0.2\,\eV$ . Note the similarity in
the scales of $a_{11}$ and $a_{22}$.

The Dirac neutrino masses are plotted in Figure \ref{hierarchy}. We note
\figfmh
in passing the possibility of extending the hierarchies to a fourth
generation. Three of the particles would have suggestively similar
masses. It turns out that to be consistent
with weak neutral current data\cite{Weak/Analysis}, the fourth up-type
quark would have
to have a similar mass to the other three particles, in conflict with
the existing hierarchy.

The probabilities for $\nu_\tau$ appearance experiments are
\begin{equation}
\begin{array}{rl}
P_{\mu\tau}&= \e^4 A^2 (-\ssd{34}+\ssd{35}+\ssd{36}+
\ssd{45}+\ssd{46}-\ssd{56}) \\
P_{e\tau}&= \e^6 A^2 \Big(-2(z-1)(\ssd{13}+\ssd{14}) -
2z(z-1)(\ssd{15}+\ssd{16})\\
&+z(\ssd{35}+\ssd{36}+\ssd{45}+\ssd{46}) - \ssd{34} + z^2\ssd{56}\Big)
\end{array}
\end{equation}
Each expression contains terms where $\Delta_{ij}$ will be large for any
conceivable experiment. For those terms, $\sin^2{\frac{\Delta_{ij}}{2}}$ will
average to $1/2$ over the finite $E$ and $L$ resolution of an experiment.
Thus we may estimate the probabilities as $P_{\mu\tau} \sim \e^4$ and
$P_{e\tau} \sim \e^6$. Currently the best limits for large $\dms$ are
$P_{\mu\tau}\lesssim0.002$ and $P_{e\tau}\lesssim0.073$ from the E531
experiment at Fermilab\cite{E531}. $P_{\mu\tau}$ gets its scale from
$V^l_{23}\sim \e^2$. Thus, while the parametrization (\ref{Wolf}) is
viable, maximal hierarchical mixing with $V^l_{23}\sim \e$ is ruled out
for any scenario with $\dms_{\mu\tau}\gtrsim10\,\eV^2$. Maximal hierarchical
mixing is not constrained in the $e$-$\tau$ channel.

Upcoming experiments may be able to distinguish this scenario from the other
possibilities in the next few years. The prediction
that the atmospheric deficit is caused by $\mu-s$ oscillations with maximal
mixing is unique to this scenario.
The mixing angle could be pinned down with further atmospheric neutrino data.
Chooz\cite{Chooz} and San Onofre\cite{San/Onofre}
can eliminate the possibility that the atmospheric deficit
is $\mu$-$e$ by directly measuring $\nu_e$ disappearance probabilities.
And ICARUS\cite{ICARUS} and MINOS\cite{MINOS} might be sufficient to rule
out $\mu$-$\tau$.
This would leave $\mu$-$s$ as the only alternative. An MeV $\nu_\tau$
is then almost inevitable to save BBN since there would be at least four
active neutrino flavors at the time of nucleosynthesis. Observations
of the solar neutrino
spectrum can firmly establish the mass splitting and mixing angle for the
neutrinos responsible, and KARMEN\cite{KARMEN}
can confirm the LSND result, which would
eliminate $\mu$-$e$ as a possibility for the solar deficit since the mass
splitting would be too large. An observation of
$\mu$-$\tau$ oscillations at CHORUS\cite{CHORUS}, NOMAD\cite{NOMAD},
or COSMOS\cite{COSMOS} would then
firmly establish $\mu$-$s$ for the solar channel.
While CHORUS and NOMAD themselves have a chance of observing $\mu$-$\tau$,
COSMOS is very likely to observe this channel, but very unlikely to see
$e$-$\tau$. 
It is clear that the next few years will be very exciting for
neutrino physics.

\section{Conclusions}

We have shown how neutrino masses appropriate
for the various oscillation data can be fit into a hierarchical mass
scenario analogous to the hierarchies in the quark and charged lepton
sectors. Small mass splittings and small mixing angles result from
the interplay of Dirac and Majorana mass terms. The scenario satisfies
all experimental and astrophysical constraints. It is
unique among proposed solutions in that the atmospheric oscillations
are $\mu$-$s$ with maximal mixing, a prediction which could be tested
experimentally in upcoming experiments.

%\bibliographystyle{prsty}
%\bibliography{hep}

\end{document}